\documentstyle[epsfig,epsf,eqsecnum,aps,amssymb,multicol]{revtex}

\begin{document}

\title{Dynamic ordering and frustration of confined vortex rows \\
studied by mode-locking experiments}

\author{N. Kokubo, R. Besseling and P. H. Kes \\}
\address{Kamerlingh Onnes Laboratory, Leiden University,P.O.Box 9504, 2300 RA Leiden, The Netherlands
}

\date{\today}
\maketitle
\begin{abstract}
The flow properties of confined vortex matter driven through
disordered mesoscopic channels are investigated by mode locking
(ML) experiments. The observed ML effects allow to trace the
evolution of both the structure and the number of confined rows
and their match to the channel width as function of magnetic
field. From a detailed analysis of the ML behavior for the case of
$3$-rows we obtain ({\it i}) the pinning frequency $f_p$, ({\it
ii}) the onset frequency $f_c$ for ML ($\propto$ ordering
velocity) and ({\it iii}) the fraction $L_{ML}/L$ of coherently
moving $3$-row regions in the channel. The field dependence of
these quantities shows that, at matching, where $L_{ML}$ is
maximum, the pinning strength is small and the ordering velocity
is low, while at mismatch, where $L_{ML}$ is small, both the
pinning force and the ordering velocity are enhanced. Further, we
find that $f_c \propto f_p^2$, consistent with the dynamic
ordering theory of Koshelev and Vinokur. The microscopic nature of
the flow and the ordering phenomena will also be discussed.

\end{abstract}
\begin{multicols}{2}
\narrowtext \noindent

\section{Introduction}
\label{secintro}

Vortex arrays (VA's) in type II superconductors are exemplary
systems to study non-equilibrium states of driven periodic media
in various pinning environments. A particularly interesting
phenomenon in this context is that of a dynamic transition from an
elastic, coherent flow state at large velocities to a plastic,
incoherent flow state at small velocities. The first theoretical
description of this issue was provided by Koshelev and Vinokur
(KV) \cite{KV} and refined in subsequent studies
\cite{Movingglass,Balents,Scheidl,Faleski} which predicted various
novel flow states, including a moving glass characterized by
elastically coupled chains oriented along the flow direction and a
moving transverse smectic with decoupled flow chains. Such
structures and the dynamic transitions between them have been
extensively studied in a number of numerical simulations
\cite{Faleski,2Dsimulation,Kolton}.

Experimentally, a diversity of flow states has been reported in
direct imaging experiments on NbSe$_2$ crystals
\cite{MarchevshkiPRL1997,PardoNature,Troyanovski}. However,
quantitatively the effect of pinning strength and/or temperature
on the ordering velocity has been studied only through dc
transport experiments \cite{BhattPRL1993,Heller,GeersPRB2001},
based on the assumption that an inflection point in dc
current-voltage ($IV$) curve (i.e. a peak in differential
resistance) marks the dynamic transition. Recently, different
explanations have been given as the reason for such inflection
point, like macroscopic coexistence of two phases
\cite{PaltielPRB2002} and a change in the self organized, large
scale morphology of vortex rivers \cite{BasslerPRB2001}. Thus, a
more direct, microscopic probe is required to study systematically
the velocity, magnetic field and temperature dependence of dynamic
ordering.

Recently, we reported on the use of mode-locking (ML) experiments
as a direct probe of ordering \cite{RutDynamicMelting}. The ML
phenomenon occurs due to coupling between, on the one hand,
collective lattice modes of frequency $f_{int}=q v_{dc}/a$, with
$q$ an integer and $a$ the lattice periodicity, which occur when a
VA moves coherently with velocity $v_{dc}$ through a pinning
potential \cite{Troyanovski,TogawaPRL2000}, and, on the other
hand, a superimposed rf-drive of frequency $f$ at an integer
fraction $1/p$ of $f_{int}$. This coupling produces steps in the
dc-transport ($IV$) curves when $v_{dc}=(p/q) f a$
\cite{Kolton,FioryHarris,Martinoli,Look,KokuboPRL02}, similar to
ML steps in sliding charge density waves (CDW's)
\cite{Gruner,Thorn1987PRB} and giant Shapiro steps in Josephson
junction arrays \cite{LeemanBenzRavindran}. However, on decreasing
the velocity $v_{dc}$ or increasing the temperature, incoherent
velocity fluctuations due to quenched and thermal disorder
suppress the collective lattice mode and reduce the width of the
ML steps. Approaching the regime of fully plastic or liquid flow,
the ML amplitude eventually vanishes
\cite{Kolton,RutDynamicMelting} and the ML frequency $f_c$ at
which this occurs provides a \emph{direct} measure of the ordering
velocity: $v_c=f_c a$.

\begin{figure}
\epsfig{file=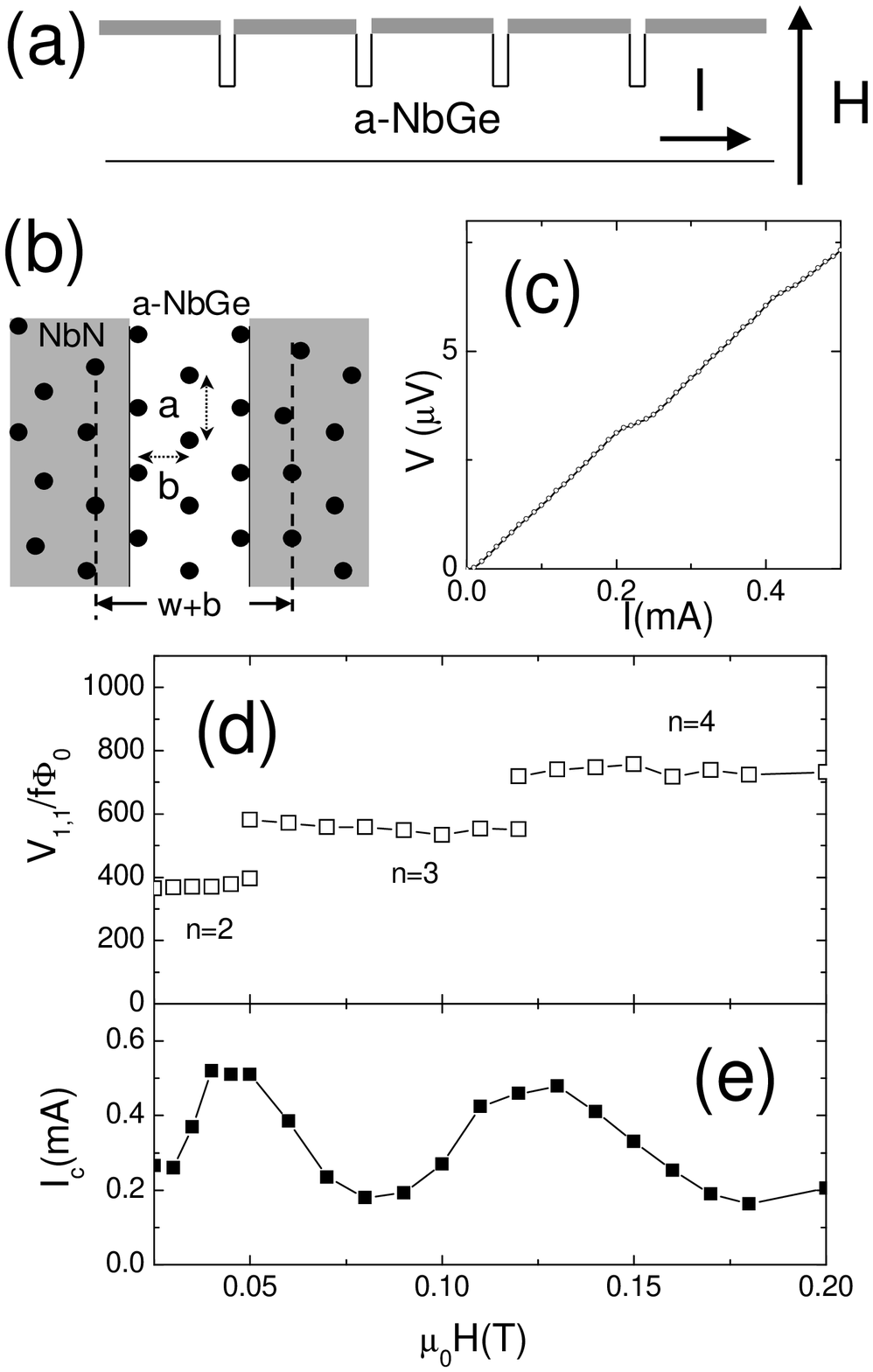, width=7cm} \vspace{0.1cm} \caption{(a)
The channel device (side view) consisting of strong pinning NbN
(dark) and weak pinning amorphous NbGe. The current and field
directions are indicated. (b) Sketch of the vortex structure
around a channel (top view). The effective channel width $w$ is
also indicated. (c) Typical ML step in a dc-$IV$ curve at $70$mT
and a superimposed rf-current of $3$MHz. (d) Normalized ML voltage
$V_{1,1}/(f \Phi_0)$ versus magnetic field. (e) Critical current
$I_c$, determined from a $1\mu$V criterion, versus field.
\label{fig1}}
\end{figure}

The particular system of our studies consists of mesoscopic flow
channels in a disordered, strong pinning environment
\cite{KokuboPRL02,PruymboomPRL}. The geometry of the samples is
sketched in Figs. \ref{fig1}(a),(b). Vortices inside the channels
are confined by strongly pinned vortices in the channel edges
(CE's). When a force is applied along the channel, the shear
interaction with these CE vortices provides the dominant pinning
mechanism impeding the channel flow. Further, since the natural
lattice (row) spacing is $a_0 \simeq 1.075\sqrt{\Phi_0/B}$
($b_0=\sqrt{3}a_0/2$), on varying the magnetic field $B$ one can
go through a series of structural transitions from $n$ to $n \pm
1$ vortex rows in the channel (typically $n \lesssim 10$). Besides
its relevance for the study of (dynamic) structural transitions of
vortex matter in quenched disorder, the physics of this system is
also closely related to layering transitions in confined fluids,
flow of colloids in mesopores and mesoscopic friction.

In Ref. \cite{KokuboPRL02} we have given a short account of the
interesting phenomena which this system displays. Firstly, due to
the structural transitions, the dc-critical current for channel
flow (yield strength) oscillates with field as shown in Fig.
\ref{fig1}(e). At a given field, the dc $IV$ (force-velocity)
curve {\it with} superimposed rf-current exhibits the ML effect,
as shown in Fig. \ref{fig1}(c). The ML condition in this case
attains a form which is particularly useful to study the
structural transitions. The voltage $V_{1,1}$ at which the
fundamental($p=q=1$) ML step occurs is given by
\cite{KokuboPRL02}:
\begin{equation} \label{modelockingV}
  V_{1,1} = f \Phi_0 n N_{ch}
\end{equation}
with $\Phi_0$ the flux quantum, $N_{ch}$ the number of channels
measured simultaneously and $n$ the {\it number of coherently
moving chains} in each channel.

As observed in Fig. \ref{fig1}(d), on changing field, $V_{1,1}$
increases as a staircase, directly reflecting the evolution of $n$
with field . A comparison with Fig. \ref{fig1}(e) shows that at
mismatch fields, i.e. where a transition $n \rightarrow n \pm 1$
occurs and $n$ and $n \pm 1$ steps may coexist, the yield strength
$\propto I_c$ is maximum, whereas around the center of the
$V_{1,1}$ plateau, where an $n$ row configuration matches the
channel width, it is minimum. As shown in Ref. \cite{KokuboPRL02},
this phenomenon is caused by positional disorder (roughness) of
the vortex configuration in the CE's: first of all, on moving away
from the matching field this disorder enhances transverse
fluctuations of vortex chains in the array, impeding the flow.
Secondly, close to mismatch, part of the (moving) $n$-chain
regions within each channel may switch to $n \pm 1$. In between
the $n$ and $n \pm 1$ regions quasi-static fault zones with
misaligned dislocations develop where the vortex trajectories are
jammed. We note that the presence of degrees of freedom transverse
to the average flow velocity, both in our system and vortex
lattices in general, forms an important difference with CDW's.
Particularly, for CDW's the 'displacement' (phase) field is a
scalar $\phi(x)$ and the velocity $\propto \partial_t \phi(x)$
represents longitudinal motion only. Our system, when compared to
regular vortex lattices, has the unique property that the
transverse response can be tuned.

In this paper we study in detail the ML step width as function of
the rf-amplitude and frequency. The experiments provide important
information on the dynamic coherence of the arrays and how this
coherency varies with mismatch and flow velocity. We focus on ML
in the regime where $n=2 \rightarrow 3 \rightarrow 4$, but our
findings are representative for other transitions with limited $n
\lesssim 10$. The paper is organized as follows. In Sec.
\ref{sectheory} we first review previous theoretical results on ML
at high frequency and then present simulation results of a one
dimensional (1D) vortex chain to show the full frequency
dependence of ML in a system with only elastic deformations. In
Sec. \ref{secexperimental} we describe the details of our sample
and the experimental procedure. The experimental results are
presented in Sec. \ref{secresults}. We find clear evidence for an
{\it ordering} frequency $f_c$ below which no ML occurs.
Furthermore, from the ML data we extract the pinning frequency
$f_p$ and the total length $L_{ML}$ of coherently moving 3-row
regions. These quantities systematically change with magnetic
field: at matching, where $L_{ML}$ is large, $f_p$ and $f_c$ are
both small, while at mismatch where $L_{ML}$ is small, both $f_p$
and $f_c$ are enhanced. We find that $f_c \propto f_p^2$,
independent of magnetic field. In Sec. \ref{secdiscus} we compare
these results with the ordering theory of Koshelev and Vinokur,
and discuss the implications for the microscopic nature of the
flow and the ordering phenomena.

\section{Theoretical Considerations}
\label{sectheory}

The velocity (or frequency) dependence of the ML step width is
particularly useful as a direct probe of dynamic ordering. At
present, the ML step width has been studied theoretically only in
the high frequency limit, where perturbation theory allows to
obtain an analytical description
\cite{SH,LarkinOvJETP74_colpinML,MartinoliPRB}.

\subsection{Equation of motion}
The $2$D displacement field $\vec{u}(\mathbf{r},\textit{t})$ of an
elastic vortex lattice driven through a pinning environment by
combined rf and dc external forces is (at $T=0$):
\begin{equation}    \label{Motion}
  \gamma \frac{\partial\vec{u}}{\partial t}= \mathbf{F}_D + \mathbf{F}_R +
  \mathbf{F}_P
\end{equation}
with $\gamma=\Phi_0 B/\rho_f$ the friction coefficient with
$\rho_f$ the flux flow resistivity. $\mathbf{F}_D$ is a driving
force per unit length consisting of dc and rf terms;
$|\mathbf{F}_D|$$=j_{dc}\Phi_0 + j_{rf}\Phi_0 \textrm{cos}(2\pi f
t)$ with $j_{dc}$ and $j_{rf}$ the dc and rf current densities,
respectively. $\mathbf{F}_R$ is the elastic restoring force given
by $(\Phi_0/B)[(c_{11}-c_{66})\nabla (\nabla \cdot \vec{u})+
c_{66}\nabla^2\vec{u}]$ with $c_{11}$ and $c_{66}$ the compression
and shear moduli, respectively \cite{KV,Faleski}. In absence of
the pinning force $\mathbf{F}_P$, the lattice is undistorted and
flows uniformly: $du/dt= \mathbf{F}_D$$/\gamma= v_{dc}+
v_{ac}\textrm{cos}(2\pi ft)$ with an ac velocity
$v_{ac}=j_{rf}\Phi_0/\gamma$, i.e. proportional to the rf-drive.

\subsection{Amplitude of the ML-interference step}
At high velocities where the friction term dominates the pinning
term in Eq.(\ref{Motion}), one can treat the pinning as a
perturbation with respect to the undisturbed rf-dc velocity
$|\mathbf{F}_D|/\gamma$. We distinguish two cases, namely a
periodic pinning potential and a random pinning potential. In case
of a periodic potential with periodicities equal to that of the
lattice, elastic deformations are absent ($\mathbf{F}_R$=0) and
the whole lattice behaves as a single particle with overdamped
dynamics in a sinusoidal potential. At large drive this case is
described analogous to a voltage biased, resistively shunted
Josephson junction \cite{Tinkham}: substituting $u = v t$ in
Eq.(\ref{Motion}) and assuming $F_P=\mu\sin(ku)$ with $\mu$ the
maximum slope of the potential and $k=2\pi/a$, one can show that
as first order correction a step anomaly appears in the dc
velocity-force characteristics at the ML condition $v_{dc}=paf$.
The current density width of the $p$th step oscillates with the
rf-drive according to
\begin{equation}    \label{Periodic}
  \Delta j_{p,1}=2j_c|J_p(v_{ac}/fa)|
\end{equation}
with $j_c=\mu/\Phi_0$ the critical current density and $J_p$ the
Bessel function of the first kind of order $p$. Note that no
subharmonic ML ($q\geq 2$) occurs in this model.

Turning to a VL in a purely random potential, the first order
perturbation correction has zero mean. The second order correction
is the lowest order of perturbation that provides the ML step.
Taking into account the lattice distortions due to the random
pinning within the elastic limit, Schmidt and Hauger \cite{SH}
showed that the ML step can appear at both harmonic and
subharmonic ML conditions $v_{dc}=(p/q)af$ and that the associated
width of the current density step is:
\begin{equation}    \label{random1}
  \Delta j_{p,q}=2 \tilde{j}_c(qk)J_p^2(qv_{ac}/fa)
\end{equation}
\begin{equation}    \label{Jcrandom}
 j_c = \sum_{q} \tilde{j}_c(qk)
\end{equation}
where $\tilde{j}_c(qk)$ is the component of the critical current
density related to the Fourier transform of the random potential
correlator at wave vector $qk$. Thus, for random pinning the ML
step width exhibits a squared Bessel-function oscillation with the
rf drive. In the following we omit the subscripts $p$ and $q$ in
the ML step width since we will discuss only the fundamental ML
phenomenon for $p=q=1$ (where $q=1$ will be justified in Sec.
\ref{secresults}). We note here that in our case $a$ can be a
frustrated lattice spacing opposed to the natural one in
\cite{SH}.

\subsection{Frequency dependence}
It is clear in Eqs.(\ref{Periodic}) and (\ref{random1}) that the
dependence on frequency and rf drive only appear in the argument
${\it z}=v_{ac}/fa$ of the Bessel functions, irrespective of the
type of pinning. Choosing ${\it z}$ by varying the rf-amplitude
such that $J_1({\it z})$ is maximum, the characteristic maximum
value of the fundamental ML width at high frequency is
\begin{equation}    \label{maximumJperiodic}
\Delta j_{max,P} = 1.16 j_c
\end{equation}
\begin{equation}    \label{maximumJrandom}
\Delta j_{max,R} =0.67 \tilde{j}_c(k)
\end{equation}
for periodic and random pinning, respectively.

These perturbational results are only applicable for frequencies
(much) above the so-called pinning frequency $f_p$. For a
sinusoidal pinning potential, $f_p$ is analogous to the
characteristic frequency of an overdamped Josephson junction and
it is given by:
\begin{equation}    \label{pinningfperiodic}
  f_p \equiv j_c\Phi_0/\gamma a.
\end{equation}
For random pinning, the fundamental ML step involves only the
$k=2\pi/a$ Fourier component of the pinning force,
$\tilde{F}_p(k)=\tilde{j}_c(k)\Phi_0$ since it is responsible for
the dynamic lattice mode excited at the washboard frequency
($f_{int}=v_{dc}/a$). In this case one can define the pinning
frequency $f_p$ \cite{fn_fpsame} such that the strength of the
friction term $\gamma af$ in Eq.(\ref{Motion}) equals
$\tilde{F}_p(k)$:
\begin{equation}    \label{pinningfrandom}
  f_p \equiv \tilde{j}_c(k)\Phi_0/\gamma a.
\end{equation}

\begin{figure}
\epsfig{file=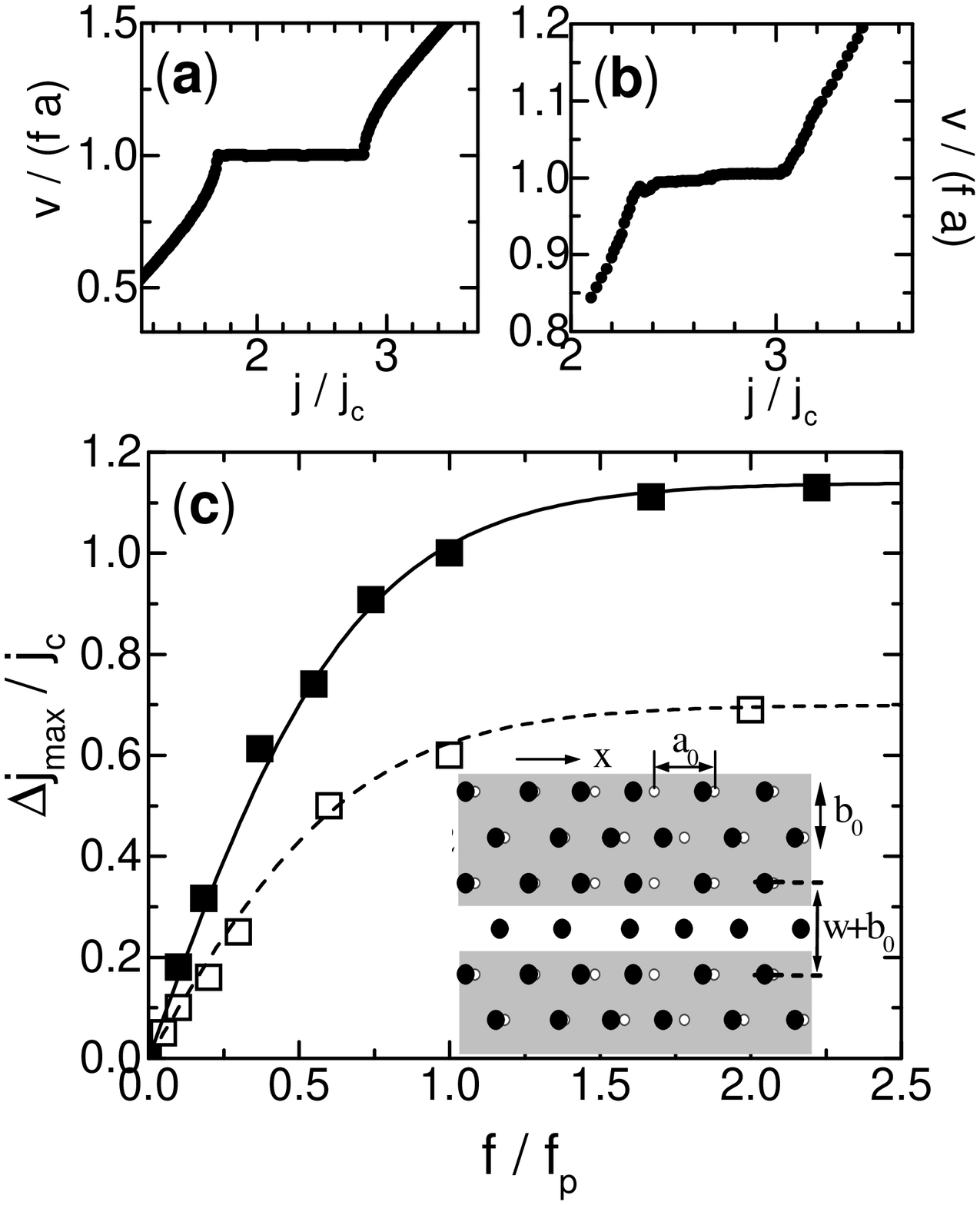, width=7.5cm} \vspace{0.1cm} \caption{(a)
Normalized force-velocity curves for simulations of a vortex chain
with an rf-drive of frequency $f=3f_p$ and amplitude
$v_{ac}/(fa)=2$ in a periodic 1D-channel; (b) same as (a) in a
disordered 1D channel. (c) Maximum ML current density step width
from simulations with rf-drive of various frequencies and
amplitudes: ($\blacksquare$) results for a channel with periodic
CE vortices; ($\square$) results for disordered 1D channels. The
curves show the empirical relation Eq.(\ref{empirical}) with
$C=1.16$ (drawn line) and $C=0.69$ (dashed line). The inset shows
the simulation geometry for a periodic CE configuration ($\circ$)
and for a disordered CE configuration ($\bullet$). \label{fig2}}
\end{figure}

Below $f_p$ no analytical result for $\Delta j_{max}$ is
available, not even for sinusoidal pinning \cite{Renne74_ML}. In
this regime numerical simulations are a useful tool to obtain the
theoretical value of the fundamental ML width
\cite{Russer1972,OctavioPRB1991}. To obtain $\Delta j_{max}$ vs.
$f$ for the case of completely elastic motion in our channel
system, we have performed molecular dynamics simulations of an
rf-dc driven 1D {\em elastic} vortex chain both in a channel with
periodically configured static vortices in the CE's and in
channels with strongly disordered CE vortex arrangements (see the
inset of Fig. \ref{fig2}(c) and \cite{RutPRL1999,BesselingEPL2003}
for more details). The channel width $w=b_0$, i.e. the (average)
spacing between the first pinned rows is $2b_0$. Vortex
interactions were modelled by the London potential with
$\lambda/a_0=1$ ($\lambda$ is the penetration depth) and the
average vortex spacing $a$ in the channel was chosen equal to that
in the CE's, $a=a_0$.

For the periodic case, the CE potential is sinusoidal and the
critical current density $j_c$ is given by its maximum slope
$j_c=\mu\Phi_0$. Simulating a chain of limited length was
sufficient since all vortices behave as a single particle. Figure
\ref{fig2} shows an example of the ML step for superimposed
rf-drive of amplitude $v_{ac}/(fa)=2$ and frequency $f\simeq
3f_p$. In Fig. \ref{fig2}(c), we summarize the numerical results
of $\Delta j_{max,P}$ versus frequency, represented by the solid
squares. Here, $\Delta j_{max,P}$ is normalized by $j_c$ and the
frequency is normalized by $f_p$ defined by
Eq.(\ref{pinningfperiodic}). At high frequency $f>f_p$, $\Delta
j_{max,P}$ saturates at a frequency independent value $\Delta
j_{s,P}/j_c \simeq 1.13$ very close to the theoretical prediction
Eq.(\ref{maximumJperiodic}) for periodic pinning. For smaller $f$,
$\Delta j_{max,P}$ starts to decrease around $f_p$ and then
vanishes linearly with $f$. The whole frequency dependence of
$\Delta j_{max,P}$ is well approximated by an empirical function
\begin{equation}    \label{empirical}
\Delta j_{max}=\Delta j_{s} \tanh(f/0.7 f_p),
\end{equation}
in which we have omitted the subscript referring to the periodic
pinning potential. Equation(\ref{empirical}) is represented by the
solid line in Fig. \ref{fig2}(c).

For the \emph{disordered} 1D channel, the CE vortices are assigned
relatively large random shifts over distances $|{\bf d}|$ with
$\sqrt{ \langle(\nabla \cdot {\bf d})^2\rangle}=0.12$ with respect
to the regular lattice configuration
\cite{BesselingEPL2003,Rutthesis}. We used a chain of 2000
vortices for which the results have become length-independent. Due
to the disorder, both the numerically obtained threshold force
$\propto j_c^R$ and the step width $\Delta j_{max,R}$ are strongly
reduced (by a factor $\sim 5$, \cite{BesselingEPL2003}) with
respect to the ordered case. An example of the simulated ML step
is shown in Fig. \ref{fig2}(b). There are two distinct differences
with the periodic case, displayed in Fig. \ref{fig2}(a): ({\it i})
the sharp corners disappear and ({\it ii}) both below and above
the ML condition the curve in (b) is essentially linear with the
same slope, with a shift at the ML condition, very similar to the
experimental result in Fig. \ref{fig1}(c). After normalization
$\Delta j_{max,R}/j_c^R$ and $f/f_p=\gamma a f/(j_c^R \Phi_0)$, we
plot the simulation results collected for various frequencies in
Fig. \ref{fig2}(c) as the open squares. The saturation value
$\Delta j_{max,R}(f \gg f_p)\equiv \Delta j_{s,R}\simeq 0.7
j_c^R$, very close to the result in the random pinning limit in
Eq.(\ref{maximumJrandom}). The whole frequency dependence is then
again well approximated by Eq.(\ref{empirical}) (dashed line in
Fig. \ref{fig2}(c)) in which we have now implicitly assumed the
subscript 'R' referring to the quantities in the random case.

It is worth mentioning that the results for the disordered channel
were insensitive to small changes in the ratio $a/a_0$. Further,
we note that both in the periodic and the disordered channel
simulations ML can be observed down to $f=0$. We believe this is a
direct consequence of the fact that vortices in the chain remain
elastically connected, because other simulations in which also
transverse degrees of freedom and plasticity are allowed, do not
show this feature \cite{BesselingchanMLtobe}.

\section{Experimental}
\label{secexperimental}

The device consists of a strong pinning layer of polycrystalline
NbN film on top of a weak pinning amorphous (\textit{a}-)
Nb$_{1-x}$Ge$_x$ film ($x \approx$ 0.3). The thickness of the NbN
and a-NbGe films are 50 and 550nm, respectively. Using reactive
ion etching with proper masking \cite{Drift}, narrow straight
channels were etched from the top (NbN) layer leaving a
300nm(=$d_{ch}$) thick (a-NbGe) bottom layer, see Fig.
\ref{fig1}(a). The width and length of each channel are 230nm and
300$\mu$m($=L$), respectively. The spacing between adjacent
channels is 10 $\mu$m. Magnetic field was applied perpendicular to
the films, inducing a vortex array as schematically shown in Fig.
\ref{fig1}(b). The transport current was applied perpendicularly
to the channel, providing a driving force parallel to the channel.

For the ML measurement, we recorded dc voltage by sweeping the dc
current with superimposed rf-current. The transmission lines for
the rf-current were terminated by matching circuits very close to
the sample. To avoid heating, both the sample and the circuits
were immersed in superfluid $^4$He. For consistency, all the data
presented in this paper were taken after field cooling in which
the magnetic field was applied above $T_c$'s of \textit{a}-NbGe
(2.68K) and NbN(11K) and the sample subsequently cooled to
$T=$1.9K, which is much lower than the vortex lattice melting
temperature $\approx 2.5K$ for the fields we studied. The ML steps
in $IV$ curves are always rounded like in Fig. \ref{fig1}(c). For
definition of the current step width $\Delta I$, we took the
derivative of the $IV$ curve and integrated over the ML peak in
the differential conductance-voltage curve with respect to the
flux-flow base line \cite{KokuboPRL02}.

\section{Results}
\label{secresults}

Our measurements were carried out in the magnetic field regime
where $3$-chain structures exist in the channels, ranging from
$45$ to $110$mT, see Fig. \ref{fig1}(d). We thus focus on the
fundamental ML step characterized by $V_{1,1}/(\Phi_0 f N_{ch})=3$
originating from {\it coherently moving} $n=3$ regions. At the
borders of our field range coexistence with $n=2$ or $n=4$ ML
steps may occur.

\begin{figure}
\epsfig{file=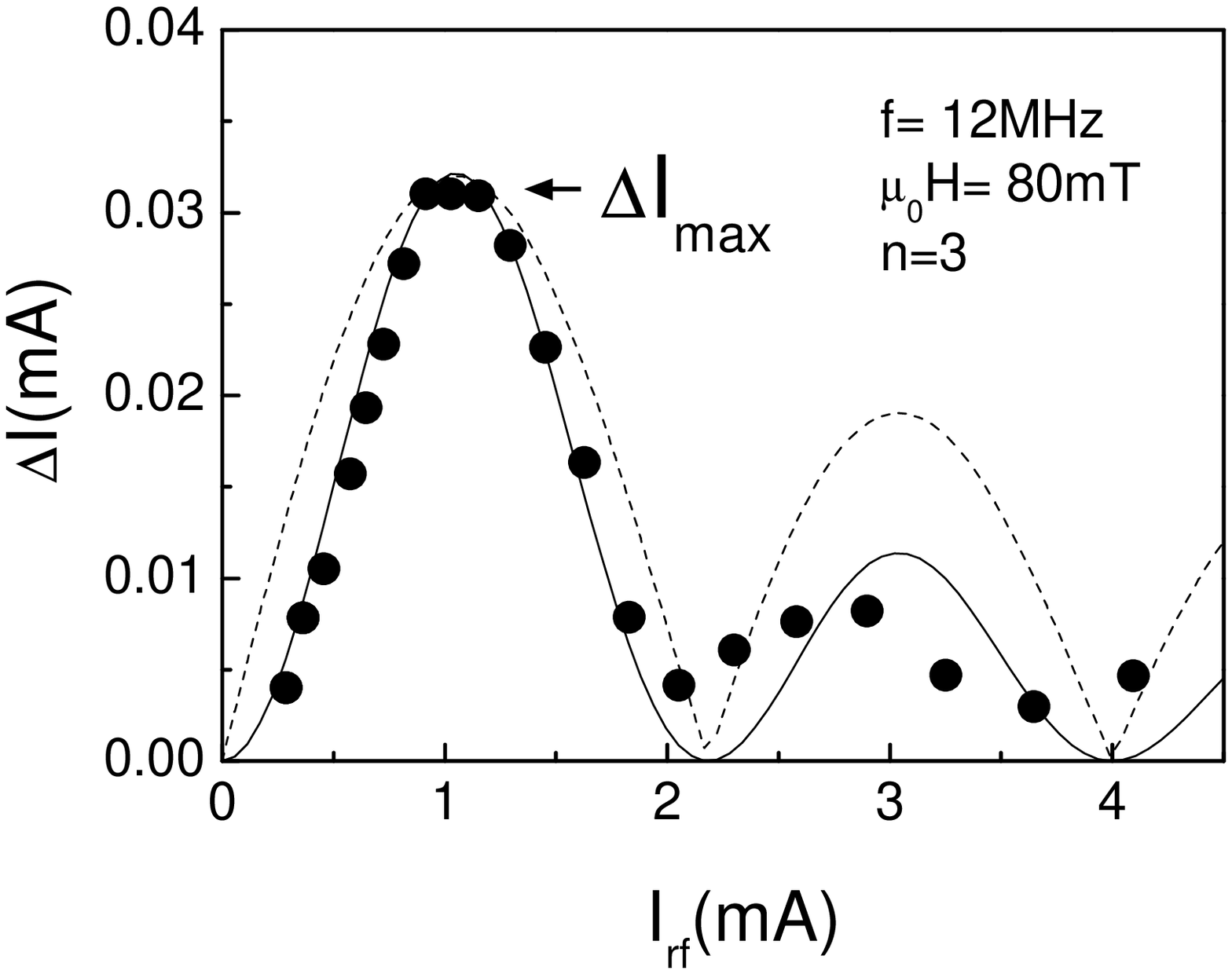, width=7.5cm} \vspace{0.1cm}
\caption{Current width $\Delta I$ of the fundamental ML step vs.
rf current $I_{rf}$ taken at 12MHz and 80mT. The maximum value
$\Delta I_{max}$ is indicated. Solid and dotted curves display
$\Delta I \propto J_1^2({\it z})$ and $\Delta I \propto |J_1({\it
z})|$, with ${\it z} \propto I_{rf}$, respectively. \label{fig3}}
\end{figure}

In Fig. \ref{fig3} we show an example of how the fundamental ML
width $\Delta I$ depends on the rf drive $I_{rf}$ for $\mu_0
H=80$mT and $f=12$MHz. $\Delta I(I_{rf})$ shows oscillatory
behavior with a maximum value $\Delta I_{max}$ in the first lobe,
as indicated in the figure. A qualitative comparison of the data
with the theoretical predictions Eq.(\ref{random1}) (solid line)
and Eq.(\ref{Periodic}) (dotted line) shows that the data follows
more closely a $J_1^2$ than a $|J_1|$ dependence. At $\Delta
I_{max}$ it is found that $I_{rf}=(1.9 \pm 0.3)I_{dc}$ over a
broad frequency range between $f_p$ and $40$ MHz (above this
frequency the experimental error in $I_{rf}$ becomes larger). This
value is in good agreement with Eq.(\ref{random1}) which has a
maximum at ${\it z}=v_{ac}/f a=v_{ac}/v_{dc} \approx
I_{rf}/I_{dc}=1.8$. It is important to note that while the values
of the rf-current might appear rather large, the actual vortex
displacements due to the rf-drive at or below the first maximum in
$J_1^2(z)$, are less than $1.8/(2\pi)\simeq 0.3$ of the lattice
spacing.

The $J_1^2(z)$ behavior shows that the pinning potential due to
the vortices in the CE acts as a random potential
(RP)\cite{Bessel}. The origin of the RP is the strong positional
disorder of the vortex lattice in the NbN edge material. This
disorder has recently been observed in scanning tunnelling
microscopy experiments on NbN films \cite{Baarle}. We further note
that we did not observe subharmonic ML steps, i.e. there were no
ML steps at $V_{ML}=3 f \Phi_0 N_{ch}/q$ with $q \geq 2$. In the
context of our RP this seems in contradiction to the results of
others \cite{FioryHarris}. But those experiments have been carried
out at relatively low fields where the RP has short range
correlations on a scale $\xi \ll a$. Fourier components $qk$ with
$q \geq 2$ are needed to describe such short range fluctuations
and therefore subharmonic ML steps are seen. In our case, the RP
is due to the CE vortices which have average spacing $a_0 \simeq
a$. Consequently the most important Fourier component describing
this RP is the $q=1$ mode, which explains why we do not see
subharmonic ML steps.

\begin{figure}
\epsfig{file=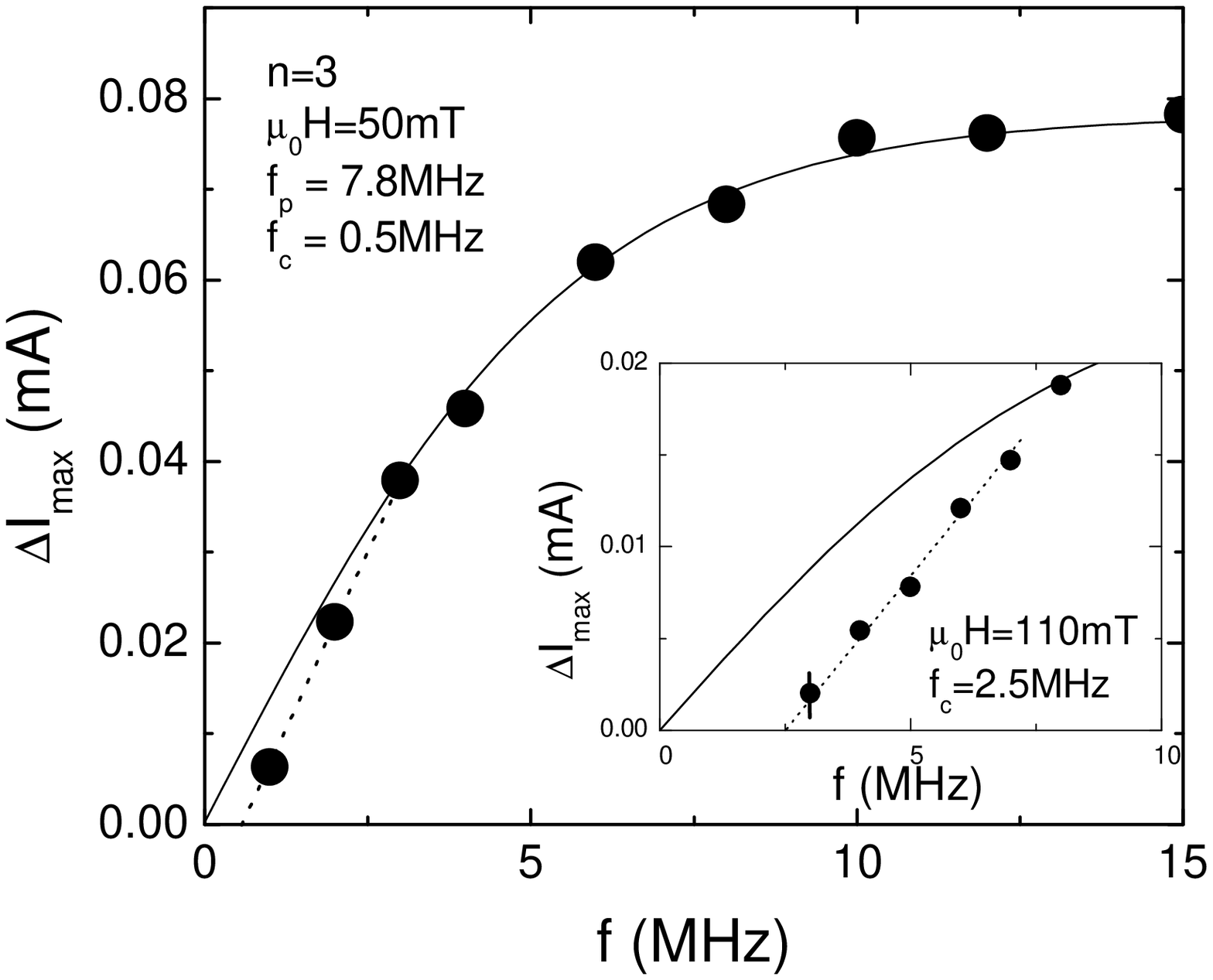, width=7.5cm} \vspace{0.1cm} \caption{The
maximum current width $\Delta I_{max}$ as a function of frequency
$f$ at $50$mT. The solid curve shows the fit according to the
empirical function Eq.(\ref{empirical}). The dotted line shows the
linear extrapolation to $\Delta I_{max}=0$ for definition of the
dynamic ordering frequency $f_c$. The inset shows the onset
behavior of $\Delta I_{max}$ and ordering frequency $f_c$ for a
field of $110$mT. The solid curve shows a fit of the high
frequency data to Eq.(\ref{empirical}). \label{fig4}}\end{figure}

Next we turn to the frequency dependence of $\Delta I_{max}$.
Figure \ref{fig4} shows $\Delta I_{max}(f)$ at $50$ mT obtained
from measurements similar to those in Fig. \ref{fig3} at various
frequencies. As in the numerical results in Fig. \ref{fig2}(c),
$\Delta I_{max}$ saturates at a value $\Delta I_{s}\simeq 78$
$\mu$A at high frequencies, while at low frequencies it decreases
monotonically with $f$. A large part of the data is well
approximated by the empirical function discussed in the previous
section, $\Delta I_{max}=\Delta I_s$tanh($f/0.7f_p)$, and we can
extract the pinning frequency $f_p$=7.8MHz as the remaining fit
parameter. However, at low frequency (i.e. small dc-velocity) the
data lie significantly below the empirical curve. This implies
that the vortex motion becomes less coherent due to the disordered
CE's. On reducing $f$, $\Delta I_{max}$ vanishes almost linearly
at a {\it finite} frequency $f_c$ determined by the intersection
between the dotted line and the $\Delta I_{max}=0$ axis. In this
regime, the rf-current for which $\Delta I$ exhibits its maximum
value, starts to saturate at a value $\sim I_c$
\cite{fn_jrfsimubelowfp}. The collapse of $\Delta I_{max}$ at
finite frequency $f_c$ is even more clearly visible in the data
taken at $110$mT shown in the inset to Fig. \ref{fig4}. Below
$f_c$ no ML step appears at \emph{any rf drive}, indicating the
complete absence of coherent $3$-row motion.

\begin{figure}
\epsfig{file=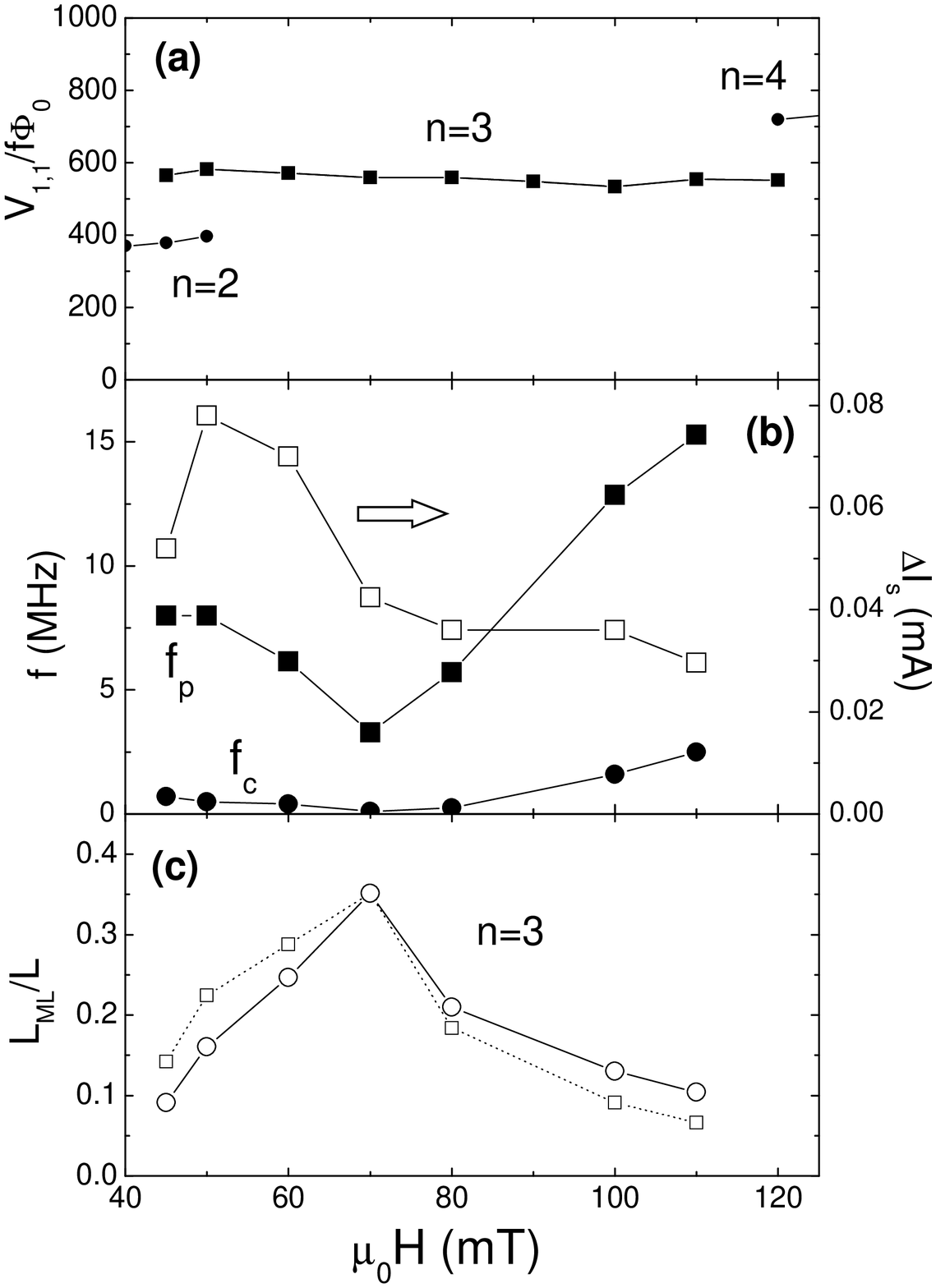, width=7.5cm} \vspace{0.1cm} \caption{(a)
$V_{1,1}/f\Phi_0$ versus field. (b) The pinning frequency $f_p$
($\blacksquare$), the dynamic ordering frequency $f_c$
($\bullet$), and the saturation value of the maximum current width
$\Delta I_{s}$ ($\square$) for $n=3$ as a function of field. The
data are obtained from fits to the measured $\Delta I_{max}(f)$.
(c) The coherently moving fraction $L_{ML}/L$ of $n=3$ regions vs.
field, where $L_{ML}$ is determined from Eq.(\ref{L}), with
$a=a_0(B)$ ($\square$) and $a=a_M B_M/B$ ($\circ$), where the
subscript 'M' refers to the matching field, see text. All lines
are guides to the eye. \label{fig5}}
\end{figure}

The above fitting analysis was performed on $\Delta I_{max}(f)$
data taken at various fields in the $n=3$ field regime and we
extracted $f_p$, $\Delta I_{s}$ and the dynamic ordering frequency
$f_c$. We first discuss the results of $f_p$ and $\Delta I_s$ as a
function of magnetic field, shown in Fig. \ref{fig5}(b). As
observed, $f_p$ has a minimum at $70$ mT, somewhat below the
middle of the plateau in $V_{1,1}$ in Fig. \ref{fig5}(a), and it
increases on approaching either end of the plateau. Hence, the
associated pinning current density in the coherent $3$-chain
regions $\tilde{j}_c(k)$ is small at $70$ mT and increases away
from $70$ mT. In fact, the value of $\tilde{j}_c(k)$ as determined
from $f_p$ using Eq.(\ref{pinningfrandom}) agrees within $30 \%$
with $j_c$ as determined directly from dc-$IV$ curves, Fig.
\ref{fig1}(e). Meanwhile, $\Delta I_s$ exhibits a field dependence
which differs considerably from that of $f_p$: it has a broad peak
around $B=50$ mT and then decreases with increasing field.
Clearly, this behavior can not be explained by simply assuming
$\Delta I_s(B)\propto j_c(B) \propto f_p(B)$.

At this point we note that theory assumes all vortices in the
channel are moving coherently. However, in our experiment only a
fraction of the vortices move coherently. Specifically, an $n$-row
region may locally break up due to the strong edge disorder or it
may coexists with $n\pm 1$-row regions (with different ML
voltages) due to mismatch \cite{KokuboPRL02}. We define the total
length of mode-locked regions with $3$ coherently moving rows as
$L_{ML}$ and the mode-locked fraction as $L_{ML}/L$. Since only
the coherent $n$-row regions contribute to $\Delta I_s$, we can
link the value of $\Delta I_s$ to that of $\Delta j_s$ by using
$\Delta I_s=L_{ML}d_{ch}\Delta j_{s}$. In this expression $\Delta
j_s$ can be obtained from the measured pinning frequency via
Eq.(\ref{pinningfrandom}) and (\ref{maximumJrandom}). Using
$\gamma=\Phi_0 B/\rho_f$ the result for $L_{ML}$ is given by
\begin{equation}    \label{L}
  L_{ML}=\frac{\Delta I_s}{f_p a} \frac{\rho_f}{0.67
 B d_{ch}}
\end{equation}
The different field dependencies of $\Delta I_s$ and $f_p$
mentioned above should thus be attributed to an additional field
dependence of $L_{ML}$.

We now evaluate $L_{ML}$ using $\rho_f$ for amorphous NbGe films
\cite{BerghuisPRB93,LarkinOvchJLTP79_fluxflow} and first assume an
equilibrium lattice spacing $a=a_0(\simeq 1.075\sqrt{\Phi_0/B})$
with $B=\mu_0 H$. Figure \ref{fig2}(c) shows $L_{ML}$ normalized
by the channel length $L$ vs. field (square symbols). As observed,
the coherently moving fraction is maximum at $B \simeq 70$ mT.
This provides a clear definition of the matching field $B_M$ for
$n=3$. At $B_M$, the longitudinal spacing $a$ in the channel
should obey $a \equiv a_M=a_0$ and the row spacing $b_M=\Phi_0/B_M
a_M$ is commensurate with the effective channel width, i.e. $3b_M
= w$. However, away from the matching field the array will be
frustrated (stretched or compressed) due to the confinement. In
particular, for $B<B_M$ the lattice spacing $a>a_0$, while for
$B>B_M$, $a < a_0$. The maximum possible difference between $a$
and $a_0$ would be achieved when the row spacing $b$ would not
change with mismatch, i.e. $b(B)=b_M$ \cite{fn_rowspacing}. This
would imply $a=(B_M/B)a_M$. Inserting this relation for $a$ in
Eq.(\ref{L}), the result for $L_{ML}$, shown by the open circles,
is slightly modified but shows essentially the same behavior as
our first analysis: upon increasing the frustration, which we
define as $|1-(B/B_M)|$, the spatial extent of regions with $3$
coherently moving rows shrinks progressively. An additional
analysis of the $n=2$ ML steps which occur at lower fields ($B
\simeq 50$mT, where the transition $n=3 \rightarrow 2$ takes
place), shows consistently that the spatial extent of the $2$ row
ML regions increases upon further decreasing field.

\begin{figure}
\epsfig{file=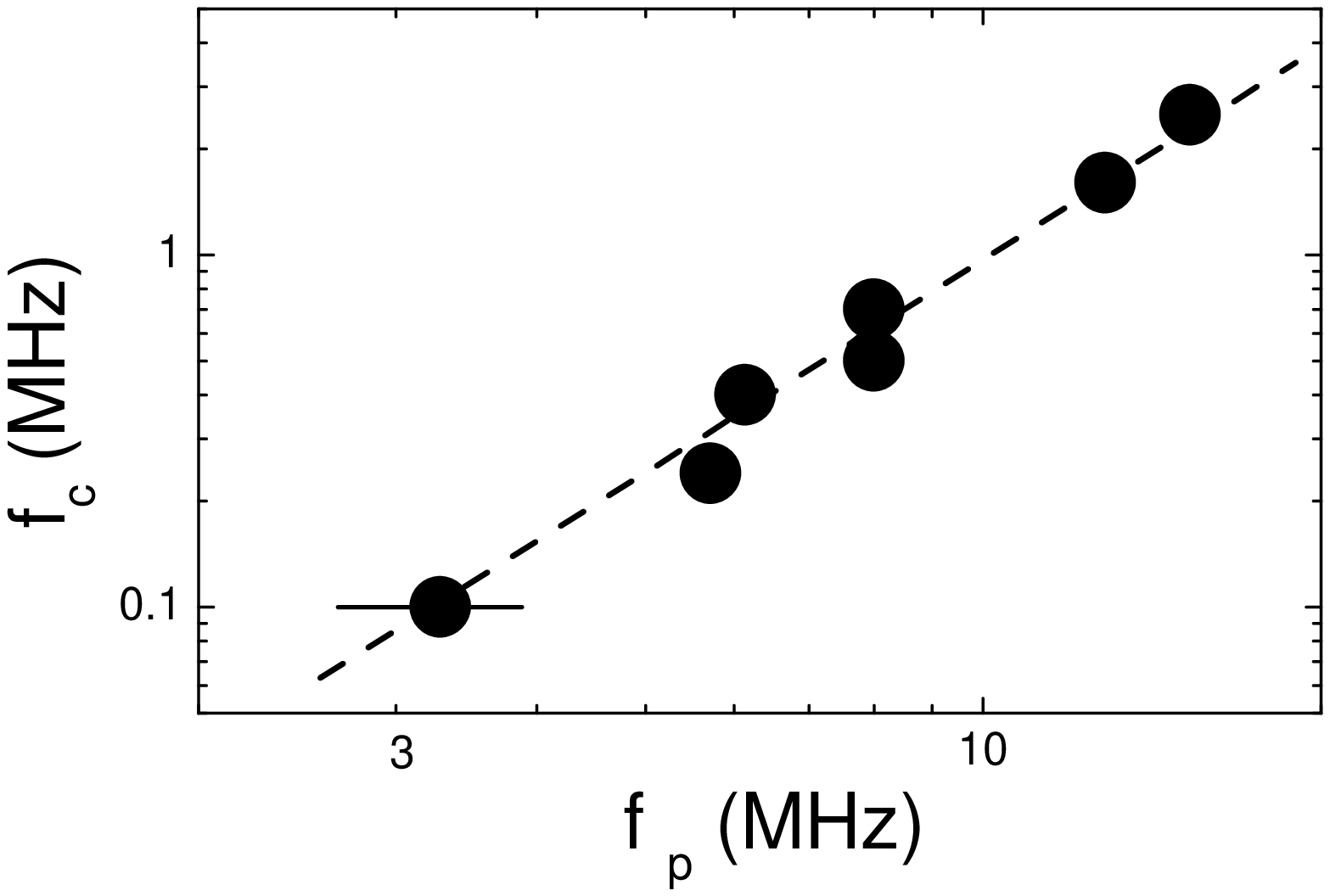, width=7.5cm} \vspace{0.1cm}
\caption{Dynamic ordering frequency $f_c$ versus pinning frequency
$f_p$. Dotted line: $f_c= \tau f_p^2$ with $\tau \simeq 1 \cdot
10^{-8}$ s. \label{fig6}}
\end{figure}

Finally we discuss the behavior of the ordering frequency $f_c$.
As shown in Fig. \ref{fig5}(b), $f_c(B)$, denoted by ($\bullet$),
exhibits a minimum at the matching field and systematically
increases with mismatch. Similarly to the decay of $L_{ML}$ with
increasing frustration, this shows that a larger mismatch
progressively induces more disorder. We also find that, when the
field is reduced below $B\simeq 50$ mT, where a $2$-row
configuration first appears, the ordering frequency $f_c^{n=2}$
for the emergence of the $n=2$ ML effect decreases.

Qualitatively, the behavior of $f_c$ is similar to that of the
pinning frequency $f_p$ (or $I_c$). Both are important quantities
characterizing the random pinning of a system and are not
independent, as follows from a double logarithmic plot of $f_c$ as
a function of $f_p$, shown in Fig. \ref{fig6}. The data are well
described by the relation $f_c = \tau f_p^2$ with $\tau \simeq 1
\cdot 10^{-8}$ s, represented by the dashed line. A more detailed
fit of the data using a power law relation $f_c \propto
f_p^\alpha$ yields an exponent $\alpha = 2.1 \pm 0.1$.

\section{Analysis and discussion of dynamic ordering and depinning threshold behavior}
\label{secdiscus}

\subsection{Comparison with the KV theory}

For a proper discussion of the above results, we first shortly
describe the phenomenological ordering theory of Koshelev and
Vinokur (KV) \cite{KV}. In their study of a 2D vortex system with
strong random bulk pinning, they found that the shaking action due
to motion through the pinning potential can be represented by a
'shaking temperature' $T_{sh}$ which decreases with velocity as
$T_{sh} \propto 1/v$. The dynamic ordering transition occurs when
the effective temperature $T+T_{sh}$ is reduced below the
equilibrium melting temperature $T_m$, i.e. when the velocity
exceeds $v_c \propto 1/(T_m-T)$. In later work
\cite{Scheidl,KoltonTeff} it was shown that the shaking
temperature refers to (bond) fluctuations transverse to the
velocity and that the associated ordering at $v_c$ corresponds to
so called transverse freezing, where inter-chain excursions
(permeation modes) are strongly suppressed. Within the KV theory
we can express the ordering frequency $f_c=v_c/a$ as:
\begin{equation}
f_c=\sqrt{3/2 \pi}\frac{\gamma_u \rho_f} {\Phi_0^2 a^2 d_{ch}
k_B(T_{m}-T)}, \label{ordfreq}
\end{equation}
with $\gamma_u$ the mean squared $2D$ pinning energy multiplied by
the area of a pin, $\rho_f$ the flux flow resistivity, $a$ the
lattice spacing and $d_{ch}$ the film thickness.

In our channel system the ordering can also be described as
transverse freezing. We observed this in simulations (e.g. for
$w/b_0\simeq 3$) as a suppression of the inter-chain excursions in
parts of the channel at sufficient velocity
\cite{BesselingchanMLtobe}. However, different from the $2$D
system considered by KV, these inter-chain excursions and the
associated shaking temperature now arise from the random
interaction with the disordered vortices in the CE's and a
modification of Eq.(\ref{ordfreq}) is required. Thinking in terms
of bond fluctuations or a Lindemann criterion, as in
\cite{Scheidl}, it is clear that it is the short wavelength $\sim
a_0$ disorder component in the potential due to quenched vortex
displacements ${\bf d}$ in the CE which is relevant for the
'shaking temperature'. This component acts only in a range $\sim
a_0/2$ from the first pinned row \cite{BesselingchanMLtobe},
therefore shaking of the outer rows should govern the transverse
freezing. A rough estimate within London theory yields an
rms-amplitude of the random stress near the edge $\sim
\varepsilon_{ce} c_{66}$ with $c_{66}$ the shear modulus and
$\varepsilon_{ce} \simeq (\sqrt{\langle|{\bf
d}|^2\rangle}/a_0)/(\pi \sqrt{3})$ representing the random strain
\cite{RutDynamicMelting,BesselingchanMLtobe}. Taking the
longitudinal range of a pin also to be $a_0/2$, we replace the
parameter $\gamma_u$ in Eq.(\ref{ordfreq}) by $\gamma_{ce}$,
resulting in:
\begin{equation} \label{gamma}
  \gamma_{ce} \simeq (\varepsilon_{ce} c_{66} a_0b_0 d_{ch})^2(\frac{a_0}{2})^2.
\end{equation}

\begin{figure}
\epsfig{file=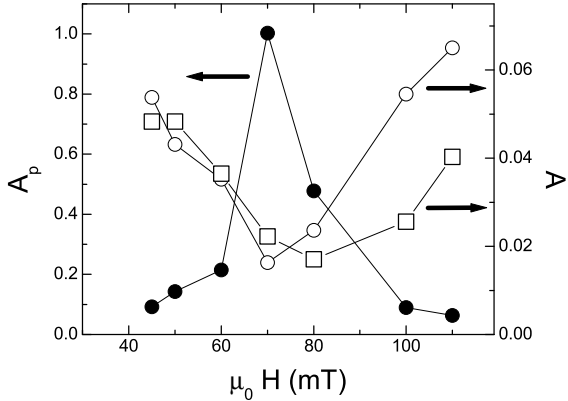, width=7.5cm} \vspace{0.1cm}
\caption{($\bullet$) The parameter $A_p$ as determined from  $f_c$
in Fig. \ref{fig5}(b) and Eq.(\ref{ordfreqmod}) using
$\varepsilon_{ce}=0.025$. ($\square$) The parameter $A$ describing
the pinning strength determined directly from
Eq.(\ref{phenomenologicaljc}) and the measured critical current
density. ($\circ$) $A$ determined from Eq.(\ref{tau}) using the
field dependence of $A_p$ and $\tau=10^{-8}$ s.} \label{fig7}
\end{figure}

Further it is important to realize the following: the energy scale
$k_B T_m$ in Eq.(\ref{ordfreq}) should be regarded as the energy
for creation of the dislocation pairs that are required for
plastic motion i.e. $k_B T_m \rightarrow k_B T_p \simeq
c_{66}a_0^2d_{ch}/(2\pi)$, with $c_{66}$ evaluated at the field
and temperature of the measurement
\cite{VinokurKesPhysC90_defects}. For our temperature and fields,
this energy $k_B T_p$ is two orders of magnitude larger than the
thermal energy $k_B T$ which we can therefore neglect in
Eq.(\ref{ordfreq}). Hence, the random shaking ($\propto 1/v$) in
our case essentially represents 'cold working' of the moving
structure. We also anticipate that the energy $k_B T_p$ should
depend on the matching condition since a reduction of this energy
eventually drives the transition to $n=2$ or $n=4$ rows. Therefore
we add a mismatch dependent factor $A_p$: $k_BT_p=A_p
c_{66}a_0^2d_{ch}/(2\pi)$, where $A_p$ is assumed to be $1$ at
matching. Taking into account these changes, Eq.(\ref{ordfreq})
becomes
\begin{equation}
f_c \simeq \frac{ \sqrt{\pi} \varepsilon_{ce}^2 c_{66}
\rho_f}{2A_p \Phi_0 B} \label{ordfreqmod}
\end{equation}
Using the experimental parameters with $c_{66}=\Phi_0 B/(16 \pi
\mu_0 \lambda^2)$ and $\lambda(1.9$ K$)\simeq 1.1$ $\mu$m, we
obtain the value of $\varepsilon_{ce}$ from the value of $f_c$ at
matching: $\varepsilon_{ce}\simeq 0.025$, i.e. $\sqrt{\langle|{\bf
d}|^2\rangle}/a_0\simeq 0.13$. This is in very reasonable
agreement with our estimate ($\sqrt{\langle|{\bf
d}|^2\rangle}/a_0\simeq 0.10$) near the melting temperature in
\cite{RutDynamicMelting}. From Eq.(\ref{ordfreqmod}) and the $f_c$
data in Fig. \ref{fig5}(b) we can also extract the field
dependence of $A_p$ characterizing the reduction of the defect
creation energy. The result is plotted in Fig. \ref{fig7}, showing
that close to mismatch $A_p$ has decreased by an order of
magnitude.

Next, we turn to the relation between $f_c$ and $f_p$. The
dc-critical current density can be described phenomenologically by
\cite{PruymboomPRL}:
\begin{equation} \label{phenomenologicaljc}
  j_c = 2A c_{66}/(Bw),
\end{equation}
where in our case $A$ varies from  $A\simeq 0.015$ at matching to
$A \simeq 0.04-0.05$ at mismatch, see the open squares in Fig.
\ref{fig7}. Further, when we combine Eqs.(\ref{ordfreq}),
(\ref{gamma}), Eq.(\ref{phenomenologicaljc}) with
Eq.(\ref{pinningfrandom}) and use $\gamma=\Phi_0 B/\rho_f$, we
obtain the quadratic relation $f_c =v_c/a= \tau f_p^2$ observed
experimentally in Fig. \ref{fig6} with the timescale $\tau$ given
by:
\begin{equation} \label{tau}
 \tau \simeq 0.5(\varepsilon_{ce}/A)^2 \frac{(w B)^2}{A_p c_{66} \rho_f}.
\end{equation}
Since $\rho_f,c_{66} \propto B$ and experimentally we found that
$\tau\simeq 10^{-8}$ s, independent of field, this implies that $A
\propto \sqrt{1/(\tau A_p)}$. Using this relation with the field
dependent value of $A_p$, we obtain $A$ as shown by the open
circles in Fig. \ref{fig7}. The minima in both data are slightly
shifted and deviations are seen for $B>B_m$, but given the
approximations made, the overall agreement is still reasonable. An
important physical implication of the relation $A \propto
\sqrt{1/(A_p)}$ is that the increase in pinning strength away from
matching is directly related to the reduction of the defect
creation energy. In other words, the rise in $A$ reflects an
effective {\it softening} of the array in the channel upon
increasing mismatch. Through this softening it is able to better
adjust to the random CE pinning potential, very much analogous to
the mechanism responsible for the peak-effect in ordinary
superconductors \cite{Pippard}. In our case a more detailed
picture of the softening mechanism is possible. We already
mentioned that for $B>B_M$, the chains are longitudinally
compressed ($a<a_0$). This will facilitate deviations in the
transverse direction and reduce the energy to (dynamically) create
interstitials between rows. For $B<B_M$, the chains are stretched,
i.e. $a>a_0$. In this case the energy for a chain to accept
vortices from a neighboring chain is lowered. This in turn
facilitates a configurational change in which the array can better
adopt to the CE potential. An additional mechanism for the rise in
$A$ for $B<B_m$ is that, due to the mismatch, the outer chains
will be pushed towards the CE, leading to an enhanced influence of
the CE potential.

To conclude this section, we want to mention that the relation
$f_c \propto f_p^2$ is not specific to our channels, but can be
obtained as a general result from the KV theory in the strong
pinning limit.

\subsection{dc versus dc+rf driven state}

So far we have tacitly assumed that the flow behavior obtained
from our rf-dc measurements simply reflects that of the dc-driven
structure. We now discuss to what extent the additional rf-current
itself influences the behavior. Recent measurements of the
rf-impedance \cite{KokuboM2S2003}, which is a sensitive probe of
ML at small rf-currents, have shown that on approaching the
dc-driven state, i.e. when $I_{rf} \rightarrow 0$, the voltage
broadening $\delta V_{1,1}$ of the fundamental ML step diverges.
Since $\delta V_{1,1} \propto \delta f_{int}$, this broadening
reflects fluctuations in the washboard frequency and, via $f=v/a$,
fluctuations in the velocity and in the longitudinal lattice
spacing $a$ \cite{fndeltaV}. Correspondingly, the divergence of
$\delta V_{1,1}$ implies that the dc-driven state lacks temporal
coherence. We also did not observe any narrow band noise in the
voltage spectrum of the dc-driven state (even for only $30$
channels at large dc-drive). At the same time, $V_{1,1}\propto n$
remains constant for $I_{rf}\rightarrow 0$, from which we conclude
that the dc-state still exhibits local regions organized in $3$
moving chains. Thus, at sufficient velocity the dc-state would
correspond to temporally incoherent, confined smectic regions
\cite{Balents} of finite (mismatch dependent) length, with
liquid-like intra chain order and residual inter-chain excursions.

In presence of rf-current the fluctuations are strongly reduced,
as also observed in experiments on CDW's \cite{BhattPRL1987}. In
simulations we observed that the suppression of the inter-chain
excursions plays an important role in this process, causing {\it
transversely frozen} regions in the channel. However, the rf-dc
$IV$ curves always show incomplete ML with the same broadening
$\delta V_{1,1}$ as discussed above. This broadening is too large
to be explained by the elastic theory in \cite{SH}. Therefore, it
is either caused by residual slip in the $n$-row regions or by
remaining plastic regions with interconnecting rows, but further
experimental and numerical work is required to decide on this
issue.

Finally, we shortly return to the frequency dependence of the ML
current width $\Delta I_{max}$ in Fig. \ref{fig4}. Extending the
relation $\Delta I_{max}(f)\propto \Delta j_{max}(f) L_{ML}(f)$ to
frequencies below $f_p$ and taking the ideal $\tanh(f)$ behavior
Eq.(\ref{empirical}) for $\Delta j_{max}$, we find that the
ordering frequency $f_c$ would mark the velocity where
$L_{ML}\rightarrow 0$. Such interpretation implies that the
dynamic ordering in our disordered system is a smooth, second
order dynamic phase transition.

\section{Summary}

Using mode-locking experiments, we have investigated the dynamics
of vortex arrays confined in disordered mesoscopic channels. The
ML effect allows to trace in detail structural transitions from
$n-1 \rightarrow n \rightarrow n+1$ confined moving vortex chains
on changing field. A study of the amplitude and frequency
dependence of the ML steps and comparison to simulations of an
{\it elastic} chain provide a complete characterization of the
pinning strength, dynamic ordering velocity and coherency of the
arrays. We find that the spatial extent $L_{ML}$ of coherently
moving $n$ row regions is large at a matching field and shrinks
with increasing mismatch. At the same time both the pinning
frequency $f_p \propto j_c$ and the ordering frequency $f_c$
(proportional to the ordering velocity) increase with mismatch. We
show that $f_c \propto f_p^2$. Together with our previous
observation of a divergence of $f_c$ near the melting temperature
in \cite{RutDynamicMelting}, these results provide detailed
experimental evidence for the phenomenological ordering theory of
Koshelev and Vinokur \cite{KV}.

\section{Acknowledgements}

We like to thank A. Koshelev for stimulating discussion. This work
was supported by the 'Stiching voor Fundamenteel Onderzoek der
Materie' (FOM) and the ESF-Vortex program.


\end{multicols}


\begin{references}
\bibitem{fn_fpsame}For convenience we use the same symbol $f_p$
both for the periodic and the random pinning cases. In the
following it will be clear from the context which one  we use.
\bibitem{fn_rowspacing} In reality the row spacing $b$
away from a matching field will be in between $b_M$ and $b_0$.
Therefore the true behavior of $L_{ML}/L$ will lie between the
dashed and the full curve in Fig. \ref{fig5}(c).
\bibitem{fndeltaV}We note that $\delta V_{1,1}$ is
much too large to be attributed to a spread in $j_c$ from channel
to channel, which was further confirmed by measurements on only
$30$ channels.
\end{references}
\end{document}